\documentclass[groupedaddress,aps,pra,superscriptaddress,showpacs,twocolumn,prl]{revtex4}%
\usepackage{epsfig,dsfont,amssymb,amsmath,amsthm,amsfonts,amsbsy,mathrsfs}
\usepackage{graphicx, color}
\usepackage{epstopdf}
\usepackage{mathdots}
\usepackage{float}
\begin{document}

\title{Polygamy relations of multipartite systems\\~}
\author{Zhi-Xiang Jin}
\thanks{Corresponding author: jzxjinzhixiang@126.com}
\affiliation{School of Physics, University of Chinese Academy of Sciences, Yuquan Road 19A, Beijing 100049, China}
\author{Shao-Ming Fei}
\thanks{Corresponding author: feishm@mail.cnu.edu.cn}
\affiliation{School of Mathematical Sciences,  Capital Normal University,  Beijing 100048,  China}
\affiliation{Max-Planck-Institute for Mathematics in the Sciences, Leipzig 04103, Germany}
\author{Cong-Feng Qiao}
\thanks{Corresponding author: qiaocf@ucas.ac.cn}
\affiliation{School of Physics, University of Chinese Academy of Sciences, Yuquan Road 19A, Beijing 100049, China}
\affiliation{CAS Center for Excellence in Particle Physics, Beijing 100049, China\\ \vspace{7pt}}

\begin{abstract}

We investigate the polygamy relations of multipartite quantum states. General polygamy inequalities are given in the $\alpha$th $(\alpha\geq 2)$ power of concurrence of assistance, $\beta$th $(\beta \geq1)$  power of entanglement of assistance, and the squared convex-roof extended negativity of assistance (SCRENoA).
\end{abstract}
\pacs{xx.xx.xx, yy.yy,yy\\}
\maketitle

\section{INTRODUCTION}

Quantum entanglement is an important kind of quantum correlation, plays essential roles in quantum information processing  \cite{MAN,RPMK,FMA,KSS,HPB,HPBO,JIV,CYSG}. One of the fundamental differences between classical and quantum correlations lies on the sharability among the subsystems. Different from the classical correlation, quantum correlation cannot be freely shared. Monogamy relation is important in the sense that it gives rise to the distribution of correlation in the multipartite quantum system and has a unique feature of keeping security in quantum key distribution \cite{MP}.	

For the systems of three qubits, a kind of monogamy of bipartite quantum entanglement in concurrence \cite{coffman} can be described by Coffman-Kundu-Wootters  CKW  inequality \cite{wootters}, $\mathcal{E}_{A|BC}\geq \mathcal{E}_{AB} +\mathcal{E}_{AC}$, where $\mathcal{E}_{A|BC}$ denotes the entanglement between systems $A$ and $BC$. Whereas monogamy of entanglement shows the restricted sharability of multipartite entanglement, the distribution of entanglement, or entanglement of assistance \cite{gg}, in multipartite quantum systems was shown to have a dually monogamous  (polygamous) property.
Note that the monogamy of entanglement inequalities provide an upper bound for bipartite sharability of entanglement in a multipartite system, and the same quantity sets a lower bound for the distribution of bipartite entanglement in a multipartite system, i.e., ${E_a}_{A|BC}\leq {E_a}_{AB} +{E_a}_{AC}$ for a tripartite quantum state $\rho_{ABC}$, where ${E_a}_{A|BC}$ is the assisted entanglement \cite{gg} between $A$ and $BC$. The polygamy inequality was first obtained in terms of the tangle of assistance \cite{gg} among three-qubit systems, and it was generalized to the multiqubit system with the help of additional  entanglement measures \cite{062328,295303,jsb}. In \cite{fgj,062302,062338}, people derived a general polygamy inequality of multipartite entanglement beyond qubit based on the entanglement of assistance.

Recently, monogamy and polygamy relations of multi-qubit entanglement have been studied in terms of non-negative
power of entanglement measures and assisted entanglement measures. In \cite{JZX, jll,ZXN}, the authors have shown that the $x$th power of the entanglement of
formation (($x\geq\sqrt{2}$)) and the concurrence ($x\geq2$) satisfy multiqubit monogamy inequalities. Monogamy relations for quantum steering have also been demonstrated in \cite{hqy,mko,jk1,jk2,jk3}.
Later, polygamy inequalities were also proposed in terms of
$\alpha$th ($0\leq\alpha\leq 1$) power of square of convex-roof extended negativity (SCREN) and the entanglement of assistance \cite{j012334, 042332}. In \cite{gy2}, the authors introduced a definition of polygamy relations without inequalities. 
However, it is still not clear for the polygamy relation of the concurrence of assistance $\tau^\alpha_a$ $(\alpha\geq 2)$ and the $\beta$th $(\beta \geq1 )$ power of entanglement of assistance $E^\beta_a$ and
the SCREN of assistance (SCRENoA) $({N_{sc}^a})^\beta$. In this paper, we study the general polygamy inequalities of $\tau^\alpha_a$, $E^\beta_a$ and $({N_{sc}^a})^\beta$ with $\alpha\geq 2$ and  $\beta \geq1$, respectively.

We first recall monogamy and polygamy inequalities related to concurrence and concurrence of assistance.
Let $\mathds{H}_X$ denote a discrete finite-dimensional complex vector space associated with a quantum subsystem $X$.
For a bipartite pure state $|\psi\rangle_{AB}\in\mathds{H}_A\otimes \mathds{H}_B$, the concurrence is given by \cite{AU,PR,SA}, $C(|\psi\rangle_{AB})=\sqrt{{2\left[1-\mathrm{Tr}(\rho_A^2)\right]}}$,
where $\rho_A$ is the reduced density matrix obtained by tracing over the subsystem $B$, $\rho_A=\mathrm{Tr}_B(|\psi\rangle_{AB}\langle\psi|)$.
The concurrence for a bipartite mixed state $\rho_{AB}$ is defined by the convex roof extension,
$C(\rho_{AB})=\min_{\{p_i,|\psi_i\rangle\}}\sum_ip_iC(|\psi_i\rangle)$,
where the minimum is taken over all possible pure state decompositions of $\rho_{AB}=\sum\limits_{i}p_i|\psi_i\rangle\langle\psi_i|$,
with $p_i\geq0$, $\sum\limits_{i}p_i=1$ and $|\psi_i\rangle\in \mathds{H}_A\otimes \mathds{H}_B$.

For a tripartite state $|\psi\rangle_{ABC}$, the concurrence of assistance is defined by \cite{TFS, YCS},
$C_a(|\psi\rangle_{ABC})\equiv C_a(\rho_{AB})=\max\limits_{\{p_i,|\psi_i\rangle\}}\sum_ip_iC(|\psi_i\rangle),$
where the maximum is taken over all possible pure state decompositions of $\rho_{AB}=\mathrm{Tr}_C(|\psi\rangle_{ABC}\langle\psi|)=\sum\limits_{i}p_i|\psi_i\rangle_{AB}\langle\psi_i|.$
For pure states $\rho_{AB}=|\psi\rangle_{AB}\langle\psi|$, one has $C(|\psi\rangle_{AB})=C_a(\rho_{AB})$.

For an $N$-qubit state $\rho_{AB_1\cdots B_{N-1}}\in \mathds{H}_A\otimes \mathds{H}_{B_1}\otimes\cdots\otimes \mathds{H}_{B_{N-1}}$,
the concurrence $C(\rho_{A|B_1\cdots B_{N-1}})$ of the state $\rho_{AB_1\cdots B_{N-1}}$, viewed as a bipartite state under the partition
$A$ and $B_1,B_2,\cdots, B_{N-1}$, satisfies the Coffman-Kundu-Wootters inequality \cite{TJ,YKM},
\begin{eqnarray}\label{C2}
  C^2(\rho_{A|B_1,B_2\cdots,B_{N-1}})\geq \sum_{i=1}^{N-1}C^2(\rho_{AB_i})\ ,
\end{eqnarray}
where $\rho_{AB_i}=\mathrm{Tr}_{B_1\cdots B_{i-1}B_{i+1}\cdots B_{N-1}}(\rho_{AB_1\cdots B_{N-1}})$.
Further improved monogamy relations are presented in \cite{JZX, ZXN}.
The dual inequality in terms of the concurrence of assistance for $N$-qubit states has the form \cite{GSB},
\begin{eqnarray}\label{DCA}
 C^2(\rho_{A|B_1,B_2\cdots,B_{N-1}})\leq \sum_{i=1}^{N-1}{C_a^2}(\rho_{AB_i})\ .
\end{eqnarray}

Now, let us consider a bipartite pure state of arbitrary dimension $d_1\times d_2$, $|\phi\rangle_{AB}=\sum_{i=1}^{d_1}\sum_{k=1}^{d_2}a_{ik}|ik\rangle_{AB}$ in $C^{d_1}\otimes C^{d_2}$.
The squared concurrence of $|\phi\rangle_{AB}$ can be expressed as \cite{jpa6777}
\begin{eqnarray}\label{beyond}
C^2(|\phi\rangle_{AB})=2(1-\mathrm{Tr}(\rho_A^2))=4\sum_{i<j}^{d_1} \sum_{k<l}^{d_2}|a_{ik}a_{jl}-a_{il}a_{jk}|^2 .
\end{eqnarray}

For a mixed state $\rho_{AB}=\sum_ip_i|\phi_i\rangle_{AB}\langle\phi_i|$, its concurrence of assistance satisfies \cite{022302}
\begin{eqnarray}
C_a(\rho_{AB}) &=& \max_{\{p_i,|\phi_i\rangle\}}\sum_ip_iC(|\phi_i\rangle)\nonumber \\
&\leq& \sum_{m=1}^{D_1} \sum_{n=1}^{D_2}(\max\sum_ip_i|\langle\phi_i|(L_A^m\otimes L_B^n)|\phi_i^\ast\rangle|) \nonumber\\
&=& \sum_{m=1}^{D_1} \sum_{n=1}^{D_2}C_a((\rho_{AB})_{mn}):=\tau_a(\rho_{AB})\; ,
\end{eqnarray} 
where 
\begin{eqnarray}
D_1 &=& d_1(d_1-1)/2,~D_2 = d_2(d_2-1)/2\ ,\\
L_A^m & =& P_A^m(-|i\rangle_A\langle j|+|j\rangle_A\langle i|)P_A^m\ ,\\
L_B^n &= & P_B^n(-|k\rangle_B\langle l|+|l\rangle_B\langle k|)P_B^n
\end{eqnarray}
with $P_A^m\!=\!|i\rangle_A\langle i|+|j\rangle_A\langle j|$ and $P_B^n\!=\!|k\rangle_B\langle k|+|l\rangle_B\langle l|$   being the projectors to the subspaces spanned by $\{|i\rangle_A, |j\rangle_A\}$ and $\{|k\rangle_B, |l\rangle_B\}$, respectively. A general polygamy inequality for any multipartite pure state $|\phi\rangle_{A_1\cdots A_n}\in C^{d_1}\otimes\cdots\otimes C^{d_n}$ was established as \cite{022302},
\begin{eqnarray}\label{cb}
\tau_a^2(|\phi\rangle_{A_1|A_2\cdots A_n})\leq \sum_{i=2}^n\tau_a^2(\rho_{A_1A_i}),
\end{eqnarray}
where $\rho_{A_1A_k}$ is the reduced density matrix $|\phi\rangle_{A_1|A_2\cdots A_n}$ with respect to subsystem $A_1A_k$, $k=2,\cdots,n$.

\section {POLYGAMY RELATION for CONCURRENCE of assistance}

{[\bf Lemma 1]}. For any real numbers $x$ and $t$, $t \geq 1$, $ x \geq 1$, we have $(1+t)^x\leq 1+(2^{x}-1)t^x$.

{\sf [Proof].} Let $f(x,y)=(1+y)^x-y^x$ with $ x\geq 1,~0<y\leq 1$, $\frac{\partial f}{\partial y}=x[(1+y)^{x-1}-y^{x-1}]\geq 0$. Therefore, $f(x,y)$ is an increasing function of $y$, i.e., $f(x,y)\leq f(x,1)=2^x-1$. Set $y=\frac{1}{t},~t\geq 1$. We obtain $(1+t)^x\leq 1+(2^x-1)t^x$. Notice when $t=1$, the inequality is true. $\Box$

The following theorem provides a class of polygamy inequalities satisfied by the $\alpha$-power of $\tau_a$.
For convenience, we denote ${\tau_a}(\rho_{AB_i})={\tau_a}_{AB_i}$ the concurrence of assistance $\rho_{AB_i}$ and ${\tau_a}(\rho_{A|B_0B_1\cdots B_{N-1}})={\tau_a}_{A|B_0B_1\cdots B_{N-1}}$.

{[\bf Theorem 1]}. For any tripartite pure state $\rho_{ABC}\in H_A\otimes H_B\otimes H_C$:

$(1)$ if ${\tau_a}_{AB}\geq {\tau_a}_{AC}$, the concurrence of assistance satisfies
\begin{eqnarray}\label{th11}
{\tau_a^\alpha}_{A|BC}\leq{\tau_a^\alpha}_{AC}+ (2^{\frac{\alpha}{2}}-1){\tau_a^\alpha}_{AB}
\end{eqnarray}
for $\alpha\geq 2$.

$(2)$ if ${\tau_a}_{AB}\leq {\tau_a}_{AC}$, the concurrence of assistance satisfies
\begin{eqnarray}\label{th12}
{\tau_a^\alpha}_{A|BC}\leq{\tau_a^\alpha}_{AB}+ (2^{\frac{\alpha}{2}}-1){\tau_a^\alpha}_{AC}
\end{eqnarray}
for $\alpha\geq 2$.

{\sf [Proof].} For arbitrary tripartite pure state $\rho_{ABC}$, one has \cite{022302}, ${\tau_a^2}_{A|BC}\leq{\tau_a^2}_{AC}+{\tau_a^2}_{AB}$. If ${\tau_a}_{AB}~({\tau_a}_{AC})=0$,
the inequality (\ref{th11}) or (\ref{th12}) are true obviously. Therefore, assuming ${\tau_a}_{AB}\geq {\tau_a}_{AC}>0$, we have
\begin{eqnarray}\label{pfth1}
{\tau_a^{2x}}_{A|BC}&&\leq({\tau_a^2}_{AB}+ {\tau_a^2}_{AC})^x\nonumber\\
&&={\tau_a^{2x}}_{AC}\left(1+\frac{{\tau_a^2}_{AB}}{{\tau_a^2}_{AC}}\right)^x\nonumber\\
&&\leq{\tau_a^{2x}}_{AC}\left(1+(2^x-1)\left(\frac{{\tau_a^2}_{AB}}{{\tau_a^2}_{AC}}\right)^x\right)\nonumber\\
&&={\tau_a^{2x}}_{AC}+(2^x-1){\tau_a^{2x}}_{AB}\ ,
\end{eqnarray}
where the second inequality is true due to the inequality $(1+t)^x\leq 1+(2^{x}-1)t^x$ for $x\geq1$ and $t=\frac{{\tau_a^2}_{AB}}{{\tau_a^2}_{AC}}\geq 1$. Denote $2x=\alpha$. We obtain $\alpha\geq2$ as $x\geq1$. Then we have the inequality (\ref{th11}). If  ${\tau_a}_{AB}\leq {\tau_a}_{AC}$, Similarly we get (\ref{th12}).

{\it Example 1.} Let us consider the three-qubit state $|\psi\rangle$ in the generalized Schmidt decomposition form,
\begin{eqnarray}\label{ex1}
|\psi\rangle&=&\lambda_0|000\rangle+\lambda_1e^{i{\varphi}}|100\rangle+\lambda_2|101\rangle \nonumber\\
&&+\lambda_3|110\rangle+\lambda_4|111\rangle,
\end{eqnarray}
where $\lambda_i\geq0,~i=0,1,2,3,4$ and $\sum\limits_{i=0}\limits^4\lambda_i^2=1.$ We have
${\tau_a}_{A|BC}=2\lambda_0\sqrt{{\lambda_2^2+\lambda_3^2+\lambda_4^2}},$
${\tau_a}_{AB}=2\lambda_0\sqrt{{\lambda_2^2+\lambda_4^2}}$, and ${\tau_a}_{AC}=2\lambda_0\sqrt{{\lambda_3^2+\lambda_4^2}}$. Without loss of generality, we set $\lambda_0=\cos\theta_0,~\lambda_1=\sin\theta_0\cos\theta_1,~\lambda_2=\sin\theta_0\sin\theta_1\cos\theta_2,~\lambda_3=\sin\theta_0\sin\theta_1\sin\theta_2\cos\theta_3$, and $\lambda_4=\sin\theta_0\sin\theta_1\sin\theta_2\sin\theta_3,~\theta_i\in[0,\frac{\pi}{2}]$.

For $\lambda_3\geq\lambda_2$, i.e. ${\tau_a}_{AC}\geq {\tau_a}_{AB}$:

(a) if $\theta_2=\frac{\pi}{2}$,
\begin{eqnarray*}
&&{\tau_a^\alpha}_{A|BC}-{\tau_a^\alpha}_{AB}- (2^{\frac{\alpha}{2}}-1){\tau_a^\alpha}_{AC}\\
&&=(2\lambda_0)^\alpha\Big[(\lambda_2^2+\lambda_3^2+\lambda_4^2)^\frac{\alpha}{2}-(\lambda_2^2+\lambda_4^2)^\frac{\alpha}{2}\\
&&~~~-(2^{\frac{\alpha}{2}}-1)(\lambda_3^2+\lambda_4^2)^\frac{\alpha}{2}\Big]\\
&&=2^\alpha\cos^\alpha\theta_0\sin^\alpha\theta_0\sin^\alpha\theta_1(2-\sin^\alpha\theta_3-2^\frac{\alpha}{2})\\&&\leq 0,
\end{eqnarray*}
where $\alpha\geq2$ and the inequality is due to $\sin\theta_3\geq 0$.

(b) If $\theta_2\neq\frac{\pi}{2}$, we denote $t_1=\frac{\sin^2\theta_2}{\cos^2\theta_2}\geq1$, then we have
\begin{eqnarray*}
&&{\tau_a^\alpha}_{A|BC}-{\tau_a^\alpha}_{AB}- (2^{\frac{\alpha}{2}}-1){\tau_a^\alpha}_{AC}\\
&&=(2\lambda_0)^\alpha\Big[(\lambda_2^2+\lambda_3^2+\lambda_4^2)^\frac{\alpha}{2}-(\lambda_2^2+\lambda_4^2)^\frac{\alpha}{2}\\
&&~~~-(2^{\frac{\alpha}{2}}-1)(\lambda_3^2+\lambda_4^2)^\frac{\alpha}{2}\Big]\\
&&=2^\alpha\cos^\alpha\theta_0\sin^\alpha\theta_0\sin^\alpha\theta_1\Big[1-(\cos^2\theta_2+\sin^2\theta_2\sin^2\theta_3)^\frac{\alpha}{2}\\
&&~~~-(2^{\frac{\alpha}{2}}-1)\sin^\alpha\theta_2\Big]\\
&&\leq 2^\alpha\cos^\alpha\theta_0\sin^\alpha\theta_0\sin^\alpha\theta_1 \Big[1-\cos^\alpha\theta_2-(2^{\frac{\alpha}{2}}-1)\sin^\alpha\theta_2\Big]\\
&&=2^\alpha\cos^\alpha\theta_0\sin^\alpha\theta_0\sin^\alpha\theta_1 \Big[1-\cos^\alpha\theta_2\left(1+(2^{\frac{\alpha}{2}}-1)t_1^\frac{\alpha}{2}\right)\Big]\\
&&\leq2^\alpha\cos^\alpha\theta_0\sin^\alpha\theta_0\sin^\alpha\theta_1\Big[1-\cos^\alpha\theta_2(1+t_1)^\frac{\alpha}{2}\Big]\\
&&=2^\alpha\cos^\alpha\theta_0\sin^\alpha\theta_0\sin^\alpha\theta_1 \Big[1-\cos^\alpha\theta_2\left(1+\frac{\sin^2\theta_2}{\cos^2\theta_2}\right)^\frac{\alpha}{2}\Big]\\&&=0 \ ,
\end{eqnarray*}
where $\alpha\geq2$ and the second inequality is due to Lemma 1.

Therefore, we have ${\tau_a^\alpha}_{A|BC}\leq{\tau_a^\alpha}_{AB}+ (2^{\frac{\alpha}{2}}-1){\tau_a^\alpha}_{AC}$ for $\alpha\geq2$. 

When $\lambda_3\leq\lambda_2$, i.e. ${\tau_a}_{AC}\leq {\tau_a}_{AB}$, from similar analysis we can obtain ${\tau_a^\alpha}_{A|BC}\leq{\tau_a^\alpha}_{AC}+ (2^{\frac{\alpha}{2}}-1){\tau_a^\alpha}_{AB}$ for $\alpha\geq2$.

Specially, when $\theta_2=\frac{\pi}{2}, \alpha=2, \theta_3=0$, i.e. $|\psi\rangle=\cos\theta_0|000\rangle+\sin\theta_0\cos\theta_1e^{i{\varphi}}|100\rangle
+\sin\theta_0\sin\theta_1|110\rangle,$ the inequality in (\ref{th11}) is saturated. Generalizing the conclusion in Theorem 1 to $N$ partite case, we have the following result.

{[\bf Theorem 2]}. For any multipartite pure state $\rho_{AB_0\cdots B_{N-1}}$, if
${\tau_a^2}_{AB_i}\leq \sum_{k=i+1}^{N-1}{\tau_a^2}_{AB_k}$ for $i=0, 1, \cdots, m$, and
${\tau_a^2}_{AB_j}\geq \sum_{k=j+1}^{N-1}{\tau_a^2}_{AB_k}$ for $j=m+1,\cdots,N-2$,
$\forall$ $1\leq m\leq N-3$, $N\geq 4$, we have
\begin{eqnarray}\label{thn1}
&&{\tau_a^\alpha}_{A|B_0B_1\cdots B_{N-1}}\leq \nonumber \\
&&{\tau_a^\alpha}_{AB_0}+(2^{\frac{\alpha}{2}}-1) {\tau_a^\alpha}_{AB_1}+\cdots+(2^{\frac{\alpha}{2}}-1)^{m}{\tau_a^\alpha}_{AB_m}\nonumber\\
&&+(2^{\frac{\alpha}{2}}-1)^{m+2}({\tau_a^\alpha}_{AB_{m+1}}+\cdots+{\tau_a^\alpha}_{AB_{N-2}}) \nonumber\\
&&+(2^{\frac{\alpha}{2}}-1)^{m+1}{\tau_a^\alpha}_{AB_{N-1}},
\end{eqnarray}
for $\alpha\geq 2$.

{\sf [Proof].} From the inequality (\ref{cb}) and Theorem 1, we have
\begin{eqnarray}\label{pfn1}
&&{\tau_a^\alpha}_{A|B_0B_1\cdots B_{N-1}}\nonumber\\
&&\leq {\tau_a^\alpha}_{AB_0}+(2^{\frac{\alpha}{2}}-1) \left(\sum_{i=1}^{N-1}{\tau^2_a}_{AB_i}\right)^\frac{\alpha}{2}\nonumber\\
&&\leq {\tau_a^\alpha}_{AB_0}+(2^{\frac{\alpha}{2}}-1){\tau_a^\alpha}_{AB_1}
 +(2^{\frac{\alpha}{2}}-1)^2\left(\sum_{i=2}^{N-1}{\tau^2_a}_{AB_i}\right)^\frac{\alpha}{2}\nonumber\\
&& \leq \cdots\nonumber\\
&&\leq {\tau_a^\alpha}_{AB_0}+(2^{\frac{\alpha}{2}}-1){\tau_a^\alpha}_{AB_1}+\cdots+(2^{\frac{\alpha}{2}}-1)^{m}{\tau_a^\alpha}_{AB_m}\nonumber\\
&&~~~~+(2^{\frac{\alpha}{2}}-1)^{m+1} \left(\sum_{i={m+1}}^{N-1}{\tau^2_a}_{AB_i}\right)^\frac{\alpha}{2}.
\end{eqnarray}
Similarly, as ${\tau_a^2}_{AB_j}\geq \sum_{k=j+1}^{N-1}{\tau_a^2}_{AB_k}$ for $j=m+1,\cdots,N-2$, we get
\begin{eqnarray}\label{pfn2}
&&\left(\sum_{i={m+1}}^{N-1}{\tau^2_a}_{AB_i}\right)^\frac{\alpha}{2} \nonumber\\
&&\leq (2^{\frac{\alpha}{2}}-1){\tau_a^\alpha}_{AB_{m+1}}+\left(\sum_{i={m+2}}^{N-1}{\tau^2_a}_{AB_i}\right)^\frac{\alpha}{2}\nonumber\\
&&\leq (2^{\frac{\alpha}{2}}-1)({\tau_a^\alpha}_{AB_{m+1}}+\cdots+{\tau_a^\alpha}_{AB_{N-2}})\nonumber\\
&&~~~~+{\tau_a^\alpha}_{AB_{N-1}}.
\end{eqnarray}
Combining (\ref{pfn1}) and (\ref{pfn2}), we have Theorem 2. $\Box$

In Theorem 2, if ${\tau_a^2}_{AB_i}\leq\sum_{j=i+1}^{N-1} {\tau_a^2}_{AB_j}$ for all $i=0,1,\cdots N-2$, then we have the following conclusion:

{\bf [Theorem 3]}.  For any multipartite pure state $\rho_{AB_0\cdots B_{N-1}}$, if
${\tau_a^2}_{AB_i}\leq\sum_{j=i+1}^{N-1} {\tau_a^2}_{AB_j}$
for all $i=0,1,\cdots N-2$, we have
\begin{eqnarray}\label{th21}
{\tau_a^\alpha}_{A|B_0B_1\cdots B_{N-1}}\leq\sum_{j=0}^{N-1} (2^{\frac{\alpha}{2}}-1)^j{\tau_a^\alpha}_{AB_j},
\end{eqnarray}
for $\alpha\geq 2$.

{\it Example 2.} We consider again the pure state (\ref{ex1}). Setting $\lambda_{0}=\lambda_{1}=\frac{1}{2}$, $\lambda_{2}=\lambda_{3}=\lambda_{4}=\frac{\sqrt{6}}{6}$, one has ${\tau_a}_{A|BC}=\frac{\sqrt{2}}{2}$, ${\tau_a}_{AB}={\tau_a}_{AC}=\frac{\sqrt{3}}{3}$. Let $y={\tau_a^\alpha}_{AB}+(2^{\frac{\alpha}{2}}-1){\tau_a^\alpha}_{AC}-{\tau_a^\alpha}_{A|BC}=2^{\frac{\alpha}{2}}\left(\frac{\sqrt{3}}{3}\right)^\alpha$, $\alpha\geq 2$,
be the residual concurrence of assistance. From our results, one can see that $y\geq 0$ for $\alpha\geq 2$, which is the case that does not given in \cite{042332}, see Fig. 1.
\begin{figure}
  \centering
  \includegraphics[width=8cm]{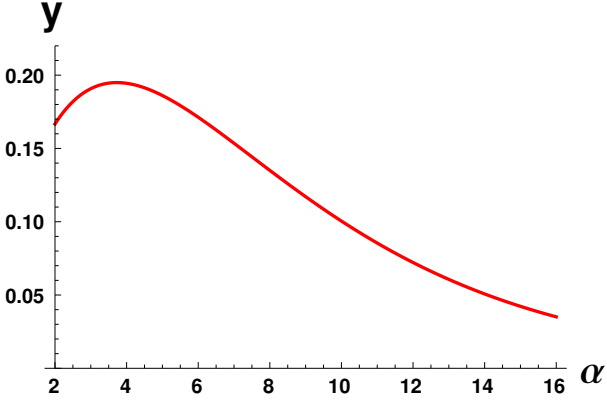}\\
  \caption{$y$ is the ``residual'' entanglement as a function of $\alpha$ with $\alpha\geq 2$.}\label{2}
\end{figure}

\section{POLYGAMY RELATIONS FOR entanglement of assistance}

For polygamy inequality beyond qubits, it was shown that the von Neumann entropy can be used to establish a polygamy inequality of tripartite quantum systems \cite{fgj}.
For any arbitrary dimensional tripartite pure state $|\psi\rangle_{ABC}$, one has $E(|\psi\rangle_{A|BC})\leq E_a(\rho_{AB})+E_a(\rho_{AC})$,
where $E(|\psi\rangle_{A|BC})=S(\rho_A)$ is the entropy of entanglement between $A$ and $BC$ in terms of the von Neumann entropy
$S(\rho)=-\mathrm{Tr}\rho\ln\rho$, and $E_a(\rho_{AB})=\max\sum_ip_iE(|\psi_i\rangle_{AB})$, with the maximization taking over all possible pure state
decompositions of $\rho_{AB}=\sum_ip_i|\psi_i\rangle_{AB}\langle\psi_i|$. Later, a general polygamy inequality
for any multipartite state $\rho_{A_1|A_2\cdots A_n}$ was established, $E_a(\rho_{A_1A_2\cdots A_n})\leq \sum_{i=2}^nE_a(\rho_{A_1A_i})$ \cite{062302}.

Recently, another class of multipartite polygamy inequalities in terms of the $\beta$th power of entanglement of assistance (EOA) has been introduced \cite{042332}.
For any multipartite state $\rho_{AB_0B_1\cdots B_{N-1}}$ and $0\leq\beta\leq 1$,
if ${E_a}_{AB_i}\geq \sum_{j=i+1}^{N-1}{E_a}_{AB_j}$ for $i=0,1,\cdots,N-2$, then
${E_a^\beta}_{A|B_0B_1\cdots B_{N-1}}\leq\sum_{j=0}^{N-1} \beta^j{E_a^\beta}_{AB_j}$, where ${E_a}(\rho_{AB_i})={E_a}_{AB_i}$ is the entanglement of assistance $\rho_{AB_i}$ and ${E_a}(\rho_{A|B_0B_1\cdots B_{N-1}})={E_a}_{A|B_0B_1\cdots B_{N-1}}$.
But, for $\beta\geq 1$ the polygamy relations for the $\beta$th  power of the entanglement of assistance is still not clear.

{[\bf Theorem 4]}. For any multipartite state $\rho_{AB_0\cdots B_{N-1}}$, if
${E_a}_{AB_i}\leq \sum_{k=i+1}^{N-1}{E_a}_{AB_k}$ for $i=0, 1, \cdots, m$, and
${E_a}_{AB_j}\geq \sum_{k=j+1}^{N-1}{E_a}_{AB_k}$ for $j=m+1,\cdots,N-2$,
$\forall$ $1\leq m\leq N-3$, $N\geq 4$, we have
\begin{eqnarray}\label{th41}
&&{E_a^\beta}_{A|B_0B_1\cdots B_{N-1}}\leq \nonumber \\
&&{E_a^\beta}_{AB_0}+(2^\beta-1) {E_a^\beta}_{AB_1}+\cdots+(2^\beta-1)^{m}{E_a^\beta}_{AB_m}\nonumber\\
&&+(2^\beta-1)^{m+2}({E_a^\beta}_{AB_{m+1}}+\cdots+{E_a^\beta}_{AB_{N-2}}) \nonumber\\
&&+(2^\beta-1)^{m+1}{E_a^\beta}_{AB_{N-1}},
\end{eqnarray}
for $\beta\geq 1$.

{\sf [Proof].} From Lemma 1, we have
\begin{eqnarray}\label{pf41}
&&{E_a^\beta}_{A|B_0B_1\cdots B_{N-1}}\nonumber\\
&&\leq {E_a^\beta}_{AB_0}+(2^\beta-1) \left(\sum_{i=1}^{N-1}{E_a}_{AB_i}\right)^\beta\nonumber\\
&&\leq {E_a^\beta}_{AB_0}+(2^\beta-1){E_a^\beta}_{AB_1}
 +(2^\beta-1)^2\left(\sum_{i=2}^{N-1}{E_a}_{AB_i}\right)^\beta\nonumber\\
&& \leq \cdots\nonumber\\
&&\leq {E_a^\beta}_{AB_0}+(2^\beta-1){E_a^\beta}_{AB_1}+\cdots+(2^\beta-1)^{m}{E_a^\beta}_{AB_m}\nonumber\\
&&~~~~+(2^\beta-1)^{m+1} \left(\sum_{i={m+1}}^{N-1}{E_a}_{AB_i}\right)^\beta.
\end{eqnarray}
Similarly, as ${E_a}_{AB_j}\geq \sum_{k=j+1}^{N-1}{E_a}_{AB_k}$ for $j=m+1,\cdots,N-2$, we get
\begin{eqnarray}\label{pf42}
&&\left(\sum_{i={m+1}}^{N-1}{E_a}_{AB_i}\right)^\beta \nonumber\\
&&\leq (2^\beta-1){E_a^\beta}_{AB_{m+1}}+\left(\sum_{i={m+2}}^{N-1}{E_a}_{AB_i}\right)^\beta\nonumber\\
&&\leq (2^\beta-1)({E_a^\beta}_{AB_{m+1}}+\cdots+{E_a^\beta}_{AB_{N-2}})\nonumber\\
&&~~~~+{E_a^\beta}_{AB_{N-1}}.
\end{eqnarray}
Combining (\ref{pf41}) and (\ref{pf42}), we have Theorem 4. $\Box$

As a special case of Theorem 4, if ${E_a}_{AB_i}\leq\sum_{j=i+1}^{N-1} {E_a}_{AB_j}$ for all $i=0,1,\cdots N-2$, we have the following conclusion:

{\bf [Theorem 5]}.  For any multipartite state $\rho_{AB_0\cdots B_{N-1}}$, if
${E_a}_{AB_i}\leq\sum_{j=i+1}^{N-1} {E_a}_{AB_j}$
for all $i=0,1,\cdots N-2$, we have
\begin{eqnarray}\label{th21}
{E_a^\beta}_{A|B_0B_1\cdots B_{N-1}}\leq\sum_{j=0}^{N-1} (2^\beta-1)^j{E_a^\beta}_{AB_j},
\end{eqnarray}
for $\beta\geq 1$.

{\it Example 3.} Let consider the three-qubit $W$ state
$|W\rangle_{ABC}=\frac{1}{\sqrt{3}}(|100\rangle+|010\rangle+|001\rangle)$. We have $E_a(|W\rangle_{A|BC})=S(\rho_A)=\log_23-\frac{2}{3}$ and
$E_a(\rho_{AB})=E_a(\rho_{AC})=\frac{2}{3}$. Set $y={E_a^\beta}(\rho_{AB})+(2^\beta-1){E_a^\beta}(\rho_{AC})-E_a^\beta(|W\rangle_{A|BC})=2^\beta(\frac{2}{3})^\beta-(\log_23-\frac{2}{3})^\beta$ to be the residual
entanglement of assistance. Fig. 2 shows our polygamy inequality for $\beta\geq1$.
\begin{figure}
  \centering
  \includegraphics[width=8cm]{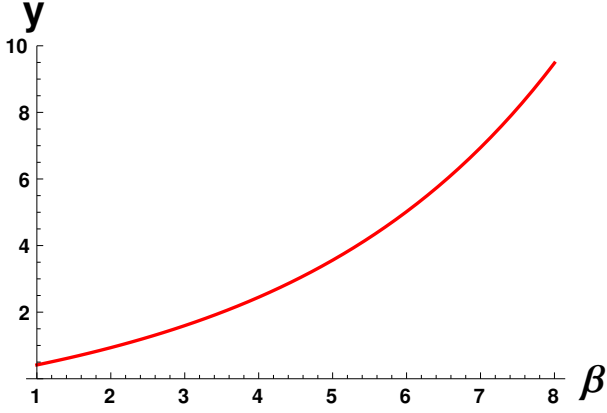}\\
  \caption{$y$ is the ``residual'' entanglement as a function of $\beta$ with $\beta\geq 1$.}\label{2}
\end{figure}

\section{POLYGAMY RELATIONS FOR SCRENoA}

Given a bipartite state $\rho_{AB}$ in $H_A\otimes H_B$, the negativity is defined by \cite{GRF},
$N(\rho_{AB})=(||\rho_{AB}^{T_A}||-1)/2$,
where $\rho_{AB}^{T_A}$ is the partially transposed $\rho_{AB}$ with respect to the subsystem $A$, $||X||$ denotes the trace norm of $X$, i.e., $||X||=\mathrm{Tr}\sqrt{XX^\dag}$.
 For the purpose of discussion, we use the following definition of negativity, $ N(\rho_{AB})=||\rho_{AB}^{T_A}||-1$.
For any bipartite pure state $|\psi\rangle_{AB}$, the negativity $ N(\rho_{AB})$ is given by
$N(|\psi\rangle_{AB})=2\sum_{i<j}\sqrt{\lambda_i\lambda_j}=(\mathrm{Tr}\sqrt{\rho_A})^2-1$,
where $\lambda_i$ are the eigenvalues for the reduced density matrix $\rho_A$ of $|\psi\rangle_{AB}$. For a mixed state $\rho_{AB}$, the square of convex-roof extended negativity (SCREN) is defined by
\begin{equation}\label{nc}
 N_{sc}(\rho_{AB})=[\mathrm{min}\sum_ip_iN(|\psi_i\rangle_{AB})]^2,
\end{equation}
where the minimum is taken over all possible pure state decompositions $\{p_i,~|\psi_i\rangle_{AB}\}$ of $\rho_{AB}$. Similar to the duality between concurrence and concurrence of assistance, we also define a dual quantity to SCREN as
\begin{equation}\label{na}
 N_{sc}^a(\rho_{AB})=[\mathrm{max}\sum_ip_iN(|\psi_i\rangle_{AB})]^2,
\end{equation}
which we refer to as the SCREN of assistance (SCRENoA), where the maximum is taken over all possible pure state decompositions $\{p_i,~|\psi_i\rangle_{AB}\}$ of $\rho_{AB}$. For convenience, we denote ${N_{sc}^a}_{AB_i}=N_{sc}^a(\rho_{AB_i})$ the SCRENoA of $\rho_{AB_i}$ and ${N_{sc}^a}_{AB_0\cdots B_{N-1}}=N_{sc}^a(|\psi\rangle_{AB_0\cdots B_{N-1}})$.

In \cite{j012334} it has been shown that ${N_{sc}^a}_{A|B_0B_1\cdots B_{N-1}}\leq \sum_{j=0}^{N-1}{N_{sc}^a}_{AB_j}$. It is further improved that for  $0\leq\beta\leq 1$, $({N_{sc}^a}_{A|B_0B_1\cdots B_{N-1}})^\beta\leq \sum_{j=0}^{N-1} \beta^j({N_{sc}^a}_{AB_j})^\beta$. But, it is still not clear whether the polygamy relation still holds for the $\beta$th $(\beta\geq 1)$ power of SCRENoA.
With a similar consideration to $\tau_{AB_0\cdots B_{N-1}}$, we have the following result of SCRENoA for $\beta\geq 1$.

{[\bf Theorem 6]}. For any multipartite state $\rho_{AB_0\cdots B_{N-1}}$, if
${N_{sc}^a}_{AB_i}\leq \sum_{k=i+1}^{N-1}{N_{sc}^a}_{AB_k}$ for $i=0, 1, \cdots, m$, and
${N_{sc}^a}_{AB_j}\geq \sum_{k=j+1}^{N-1}{N_{sc}^a}_{AB_k}$ for $j=m+1,\cdots,N-2$,
$\forall$ $1\leq m\leq N-3$, $N\geq 4$, we have
\begin{eqnarray}\label{th61}
&&({N_{sc}^a}_{A|B_0B_1\cdots B_{N-1}})^\beta\leq ({N_{sc}^a}_{AB_0})^\beta \nonumber \\
&&+(2^\beta-1) ({N_{sc}^a}_{AB_1})^\beta+\cdots+(2^\beta-1)^{m}({N_{sc}^a}_{AB_m})^\beta\nonumber\\
&&+(2^\beta-1)^{m+2}\left(({N_{sc}^a}_{AB_{m+1}})^\beta+\cdots+({N_{sc}^a}_{AB_{N-2}})^\beta\right) \nonumber\\
&&+(2^\beta-1)^{m+1}({N_{sc}^a}_{AB_{N-1}})^\beta,
\end{eqnarray}
for $\beta\geq 1$.

{\sf [Proof].} From Lemma 1, we have
\begin{eqnarray}\label{pf61}
&&({N_{sc}^a}_{A|B_0B_1\cdots B_{N-1}})^\beta\nonumber\\
&&\leq ({N_{sc}^a}_{AB_0})^\beta+(2^\beta-1) \left(\sum_{i=1}^{N-1}{N_{sc}^a}_{AB_i}\right)^\beta\nonumber\\
&&\leq ({N_{sc}^a}_{AB_0})^\beta+(2^\beta-1)({N_{sc}^a}_{AB_1})^\beta\nonumber\\
&&~~~~+(2^\beta-1)^2\left(\sum_{i=2}^{N-1}{N_{sc}^a}_{AB_i}\right)^\beta\nonumber\\
&& \leq \cdots\nonumber\\
&&\leq ({N_{sc}^a}_{AB_0})^\beta+(2^\beta-1)({N_{sc}^a}_{AB_1})^\beta+\cdots\nonumber\\
&&~~~~+(2^\beta-1)^{m}({N_{sc}^a}_{AB_m})^\beta\nonumber\\
&&~~~~+(2^\beta-1)^{m+1} \left(\sum_{i={m+1}}^{N-1}{N_{sc}^a}_{AB_i}\right)^\beta.
\end{eqnarray}
Similarly, as ${N_{sc}^a}_{AB_j}\geq \sum_{k=j+1}^{N-1}{N_{sc}^a}_{AB_k}$ for $j=m+1,\cdots,N-2$, we get
\begin{eqnarray}\label{pf62}
&&\left(\sum_{i={m+1}}^{N-1}{N_{sc}^a}_{AB_i}\right)^\beta \nonumber\\
&&\leq (2^\beta-1)({N_{sc}^a}_{AB_{m+1}})^\beta+\left(\sum_{i={m+2}}^{N-1}{N_{sc}^a}_{AB_i}\right)^\beta\nonumber\\
&&\leq (2^\beta-1)\left(({N_{sc}^a}_{AB_{m+1}})^\beta+\cdots+({N_{sc}^a}_{AB_{N-1}})^\beta\right)\nonumber\\
&&~~~~+({N_{sc}^a}_{AB_{N-1}})^\beta.
\end{eqnarray}
Combining (\ref{pf61}) and (\ref{pf62}), we have the Theorem 6. $\Box$

Particularly, the equality in (\ref{th61}) can be established for 4-qubit generlized $W$-class states $|W\rangle_{AB_1B_2B_3}=\frac{1}{2}(|1000\rangle+|0100\rangle+|0010\rangle+|0001\rangle)$, with $\beta=1$, which can be seen clearly in example 4 below Theorem 7. Specially, from Theorem 6 we have

{\bf [Theorem 7]}.  For any multipartite state $\rho_{AB_0\cdots B_{N-1}}$, if
${E_a}_{AB_i}\leq\sum_{j=i+1}^{N-1} {E_a}_{AB_j}$
for all $i=0,1,\cdots N-2$, we have
\begin{eqnarray}\label{th71}
({N_{sc}^a}_{A|B_0B_1\cdots B_{N-1}})^\beta\leq\sum_{j=0}^{N-1} (2^\beta-1)^j({N_{sc}^a}_{AB_j})^\beta
\end{eqnarray}
for $\beta\geq 1$.

{\it Example 4}. Let us consider the 4-qubit generlized $W$-class states,
\begin{eqnarray}\label{W4}
|W\rangle_{AB_1B_2B_3}=\frac{1}{2}(|1000\rangle+|0100\rangle+|0010\rangle+|0001\rangle).
\end{eqnarray}
We have ${N_{sc}^a}_{A|B_1B_2B_3}=\frac{3}{4}$, ${N_{sc}^a}_{AB_i}=\frac{1}{4}$, $i=1,2,3$. Let $y$ be the difference between the right and left hand of inequality (\ref{th71}). One has $y=[2^\beta+(2^\beta-1)^2](\frac{1}{4})^\beta-(\frac{3}{4})^\beta$; see Fig. 3.
\begin{figure}
  \centering
  \includegraphics[width=8cm]{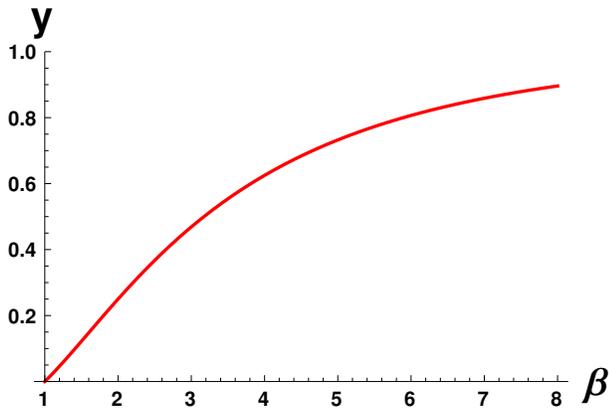}\\
  \caption{The ``residual'' entanglement $y$ as a function of $\beta$ ($\beta\geq 1$).}\label{2}
\end{figure}

\section{conclusion}

Entanglement monogamy and polygamy are fundamental properties of multipartite entangled states. We have investigated in this work the polygamy relations related to the concurrence of assistance, entanglement of assistance, and SCREN generally for multipartite states. We have found a class of polygamy inequalities of multipartite entanglement in arbitrary-dimensional quantum systems in the $\alpha$th ($\alpha\geq 2$) power of concurrence of assistance, a case that has not been studied before. Moreover, the $\beta$th power of polygamy inequalities have been obtained for the entanglement of assistance and SCRENoA for $\beta \geq 1$. The approach developed in this work is applicable to the study of monogamy properties in other quantum entanglement measures and quantum correlations.

\bigskip

\noindent{\bf Acknowledgments}\, \,

This work was supported in part by the National Natural Science Foundation of China(NSFC) under Grants 11847209; 11675113 and 11635009; Key Project of Beijing Municipal Commission of Education under No. KZ201810028042; the Ministry of Science and Technology of the Peoples' Republic of China(2015CB856703); and the Strategic Priority Research Program of the Chinese Academy of Sciences, Grant No. XDB23030100.

\end{document}